# Core-shell multi-quantum wells in ZnO / ZnMgO nanowires with high opticalefficiency at room temperature

R Thierry[1], G Perillat-Merceroz[1,2], P H Jouneau[1,2], P Ferret[1]*, G Feuillet[1]

1.- CEA, LETI, Minatec Campus, 38054, Grenoble, France

2.- CEA-INAC/UJF-Grenoble1 UMR-E, SP2M, LEMMA, Minatec Campus, 38054, Grenoble, France

Correspondingauthor : pierre.ferret@cea.fr

Nanowire-based light-emitting devices require multi-quantum well heterostructures with high room temperature optical efficiencies. We demonstrate that such efficiencies can be attained through the use of $ZnO/Zn_{(1-x)}Mg_xO$ core shell quantum well heterostructures grown by metal organic vapour phase epitaxy. Varying the barrier Mg concentration from x=0.15 to x=0.3 leads to the formation of misfit induced dislocations in the multi quantum wells. Correlatively, temperature dependant photoluminescence reveals that the radial well luminescence intensity decreases much less rapidly with increasing temperature for the lower Mg concentration. Indeed, about 54% of the 10K intensity is retained at room temperature with x=0.15, against 2% with x=0.30. Those results open the way to the realization of high optical efficiencynanowire-based light emitting diodes.

1. **Introduction**

Because of its remarkable optical properties, ZnO has attracted the attention of many groups working on light-emitting devices. ZnO nanowires are of particular interest because they can grow without any catalysts (avoiding contamination) on a variety of substrates such as sapphire,[1,2] ZnO,[3,4] GaN,[5] silicon,[5] metals,[6] graphene[7] and glass.[8] Moreover, they exhibit good structural[9] and optical properties,[10] and consequently may act as good templates to grow heterostructures. Core-shell quantum well heterostructures have been grown on nanowires because of a larger active region than in two dimensional heterostructures or 1D axial nanowire heterostructures. A single quantum well ZnO / ZnMgO core-shell heterostructurewas first demonstrated on nanowires grown by metal organic vapor phase epitaxy (MOVPE)[11]. Core-shell quantum wells were also grown by pulsed laser deposition on a low density array of vertical nanowires.[12,13] Because these devices have to work at room temperature, it is crucial to assess how the luminescence intensity is maintained from low temperature – typically 10K – to room temperature.  Although the ratio between the photoluminescence (PL) intensity at 10K over the one at room temperature (hereafternamed PL IQE) is not properly speaking equal to the internal quantum efficiency of the emitting structure, it nonetheless gives a good estimate of this figure of merit (more information on PL IQE measurements can be found for example in Ref [14]) For two-dimensional (0001)ZnO/ZnMgO-quantum well heterostructures, quantum confinement was demonstrated at low temperatures with quantum well emissions below the ZnO gap, because of the Quantum Confined Stark Effect (QCSE).[15]QCSE reduces the electron-hole wave function overlap and thereforeis detrimental for light emission efficiency.[16]In this respect, *a*-oriented non-polar quantum well heterostructures grown by molecular beam epitaxy (MBE) were shown to lead to increased optical efficiencies.[17]Interestingly, in the case the layers were grown heteroepitaxially onto *r*-plane sapphire, the PL IQE was only of the order of 1%, whereas for homoepitaxial growth on *a*-plane ZnO substrates, this ratio reached 20%, stressing the detrimental effect of extended defects (as introduced upon misfit relaxation) upon light emission at room temperature. Lately, further improvements of the PL IQE were obtained when MBE growth was carried out on *m*-plane ZnO.[18]

Along this line, since ZnO nanowires grow defect-free along the c direction, and have *m*-plane lateral facets[9], they can be used as templates for growing core-shell quantum well ZnO/ZnMgO heterostructures. Two groups have reported the growth of core-shell structures, but no figure was ever given for the PL IQE.[11-13]

It is the aim of this paper to demonstrate that high optical efficiencies can be obtained at room temperature from core-shell ZnO / ZnMgO multi-quantum well heterostructures. It is also shown that these optical efficiencies, as evaluated from the PL IQE, strongly depend upon the Mg concentration in the barriers. This gives strong evidence that misfit induced dislocations, which are present for higher Mg concentration, as revealed by transmission electron microscopy, lead to non-radiative recombination. Moreover, since the multi-quantum well core-shell structures are grown on well-aligned, dense arrays of nanowires, their integration into a device such as a light-emitting diode is greatly facilitated.[19]

2. **Experimental details**

The samples were grown in an industrial Aixtron MOVPE system . The reactor is equipped with a close coupled showerhead gas injector which prevents pre-reactions from occurring in the gas phase, and leads to good homogeneity and reproducibility. Moreover, the small distance between injection and sample surface allows for abrupt and accurate composition variations. In a first step, ZnO nanowire arrays were grown at 850°C using $N_2O$ and diethylzinc (DeZn), with a $n(N_2O)/n(DeZn)$ molar ratio of about 500. $N_2$ was the carrier gas, and the pressure during growth was set at 100mbar. In a second step, the ZnMgO/ZnO radial heterostructure was grown at a lower temperature of 500°C to avoid Mg and Zn interdiffusion,[20] and with higher $n(N_2O)/n(DeZn)$ ratio (typically above 5000) to promote radial growth instead of axial growth. $MeCp_2Mg$ was used as a Mg precursor for the growth of the ZnMgO barriers. Multi-Quantum Well (MQW) heterostructures with three ZnO quantum wells and four ZnMgO barriers were grown radially onto the ZnO core. In order to study the effect of the Mg composition in the barriers on the optical properties, two samples named MQW1 and MQW2 were grown with two injected precursor molar ratio $R=n(MeCp2Mg)/n(DeZn)$, respectively R=1 and R=2. For the first sample, growth duration for the ZnO core, the 3 identical ZnMgO barriers and the 3 identical ZnO quantum wells were respectively 1500 s, 200 s and 50 s. For the second sample, growth durations were respectively 1500 s, 250 s and 75 s. Two calibration samples namedCS1 and CS2 were made to determine the Mg composition of the $Zn_{(1-x)}Mg_xO$ alloy: a single ZnMgO shell of about 100 nm was grown on the ZnO nanowire core samples with R= 1 and R = 2. Transmission electron microscopy (TEM) was carried out on a FEI-Tecnai microscope operated at 200 kV. Weak-beam TEM was used in order to image the diffraction contrast caused by structural defects, and the scanning TEM (STEM) mode with a high angle annular dark field (HAADF) detector to allow for mass contrast. Sample preparation was carried out by the cleaved edge method for the cross section views: small triangles of sapphire substrates with vertical ZnO nanowires were cleaved and glued on copper grids in order to align the nanowires perpendicularly to the electron beam.[21]Focused ion beam (FIB) was used for the plan view samples. The nanowires were dispersed on a silicon substrate and protected by a first platinum layer deposited in an evaporator. Then a thicker platinum layer was deposited in the FIB.A piece of Si with the dispersed nanowireswas extracted, glued on a copper grid, and milled until electron transparency. Nanowires were mechanically dispersed on a silicon conductive substrate in order to analyze their optical properties without any contribution from underlying pyramids, unintentionally grown during the process.[10] Spatially resolved cathodoluminescence (CL) spectroscopy was carried out on single nanowires. Measurements were performed at 5K using a 15 keV electron beam with a 50 nm spot size. Finally, photoluminescence spectroscopy was performed on a nanowire assembly from 10K to 300K with a frequency-doubled 244 nm continuous wave Ar laser coupled with a 0.55 mmonochromator equipped with a 600 $mm^{-1}$ grating. The spot size diameter was about 50μm and the excitation power was 30 μW. The incidence angle of the laser beam with the surface was about 45°.

### 3. Determination of Mg concentrations

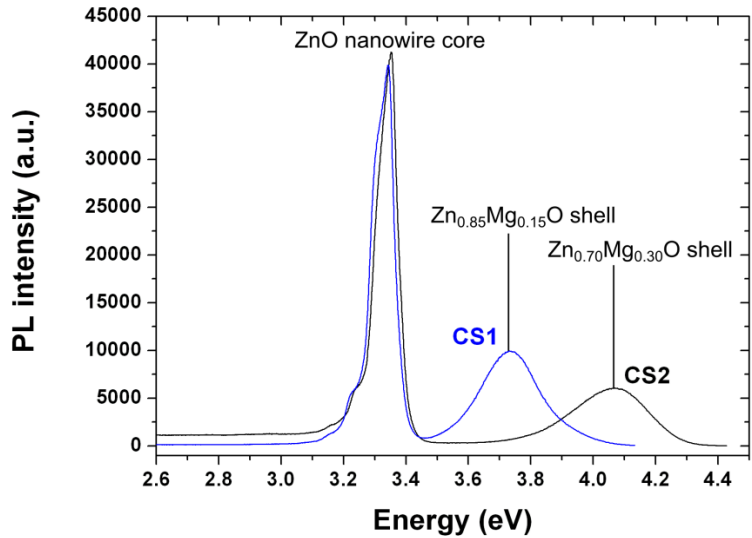

Figure 1: Room temperature PL spectra of core-shell structures with single ZnMgO shells on ZnO cores with two different Mg concentrations (CS1 with R=1, CS2 with R=2).

The $Zn_{(1-x)}Mg_xO$ alloycomposition was determined from PL measurements. PL was preferred to energy dispersive x-ray spectrometry (EDX) because it is difficult to obtain a quantitative composition in core-shell nanowires from EDX. Dedicated calibration samples with a single ZnMgO shell on a ZnO core were grown. This is necessary, since the emission from the ZnMgO barriers is not observed on samples with ZnO quantum wells, because of the highexciton diffusion length compared to the barrier thickness.[22] Figure 1 shows the PL spectra of these two samples. In addition to the ZnO excitonic emission at 3.36eV, a peak at 3.760 eV (respectively 4.080 eV) is observed for CS1 sample (respectively CS2), which is attributed to photoluminescence from the ZnMgO shell. The Mg induced blue shift in a $Zn_{(1-x)}Mg_xO$ alloy has been "calibrated" with EDX and PL results obtained from ZnMgO two-dimensional thin films (results not shown here), leading to a shift in energy of 25 meV per percent of Mg. Note that this Vegard's law is consistent with the one proposed by Koike *et al*.[23] This yields compositions of 15% and 30% for CS1 and CS2 samples, respectively.

### 4. Axial and radial quantum wells

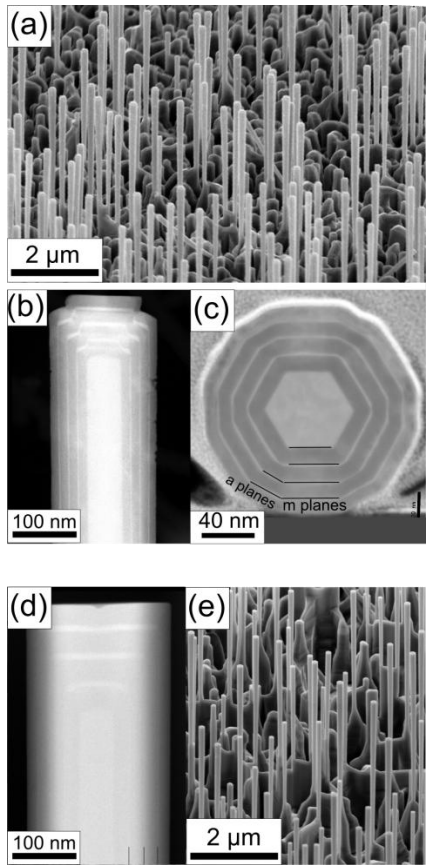

Figure 2: Images of sampleMQW2 (30% Mg): (a) 20° tilted SEM image, (b) cross section HAADF-STEM image, (c) plan view HAADF-STEM image . Images of sampleMQW1 (15% Mg): (d) HAADF-STEM image, and (e) 20°tilted SEM image.

Figure 2 (a) and (e) are SEM images of samples MQW2and MQW1. The [0001] aligned nanowires have a length in between 2 and 3µm and diameters from 150 to 250 nm. This dispersion is due to spontaneous growth: neither catalyst nor selective area pattern were used. A higher radial growth rate on the top of the nanowires is observed, as evidenced by theirinverse tapering shape.In the HAADF-STEM image of MQW2 sample on Figure 2 (b), the MQW structure is clearly visible because, in this mode, the heavier ZnO appears brighter than the lighter ZnMgO. Despite rough interfaces, STEM images reveal that the quantum wells and the barrier have mean thicknesses of respectively about 4 nm and 10 nm.The heterostructure growth also occurred in the axial direction: three axial quantum wells are visible at the nanowire top.The axial quantum well width isaround 8 nm: thegrowth rate is almost twice higher than in the radial direction.On Figure 2 (c), the STEM image of a nanowire slice from sample MQW2 is presented. The ZnO core, the four ZnMgO barriers, and the three ZnO quantum wells are visible. The sample was prepared by FIB, and a thin and very bright platinum layer appears on the top of the nanowire slice. The ZnO core has a hexagonal shape with *m*-plane facets (TEM image with the corresponding diffraction pattern not shown here, see also Ref [9]). The first ZnO quantum well is also faceted with *m*-planes, but, onincreasing the radius of the nanowire, a-plane facets appear. Figure 2 (d), is a STEM image of sample MQW1. Because of the lower Mg concentration (15%), the contrast between the ZnMgO barriers and the ZnO wells is much weaker. However, the three axial and radial ZnO quantum wells are still discernible (The position of thethree radial wells isindicated by black lines at the bottom right of Figure 2 (d)). Although the quantum well growth time was a bit shorter than for sample MQW2, the actual measured width is around 4.5 nm. This discrepancy could arise because the length of the nanowires and their diameter can vary slightly over the wafer because of local nanowire density fluctuations. As a matter of fact, on the cleaved edge observed by TEM [Figure 2 (d)], the MQW1 nanowires were larger than those observed by SEM [Figure 2 (e)].

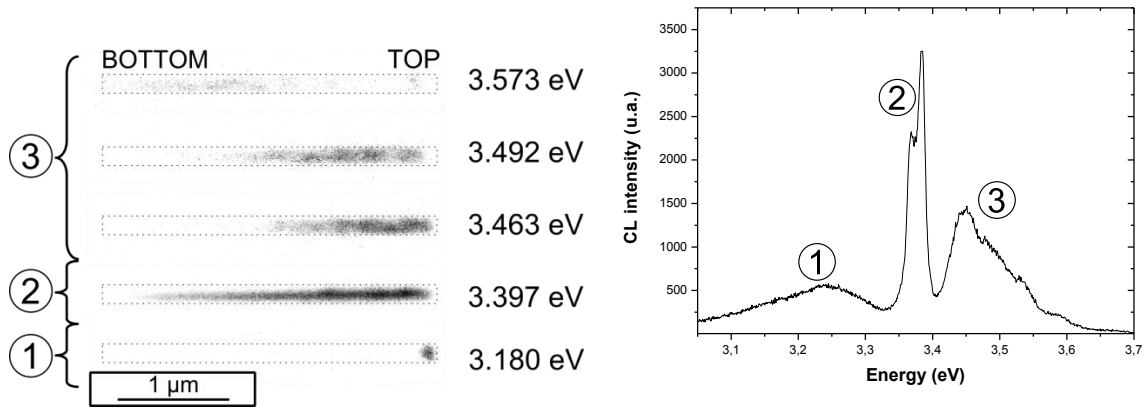

Figure 3 : Spatially resolved CL spectroscopy on a single nanowire from MQW 2 sample (Light emission appears in black). Emissions from ZnO near band edge (2), axial (1) and radial (3) quantum wells are clearly visible.

Spatially resolved CL spectroscopy results are presented on Figure 3. Three types of contributions are observed. The near band edge contributions (denoted (2) on Figure 3) are observed on the wholenanowire length and are usually attributed to neutral donor bound excitons. Aluminum coming from the substrate or hydrogen from the precursors could be the involved donors.[24] The below band gap emission (noted (1) on Figure 3) is only present on the top of the nanowire and is attributed to the axial quantum wells, as observed on the STEM images on Figure 2 (b)). This emission below the ZnO gap energy is due to the QCSE, because of the presence of a spontaneous electrical field along the c-axis in polar ZnO. Although impurities could also be the cause of an emission under the ZnO band gap, it should not be the case here. Actually, the ZnO nanowires (without any heterostructures) do not exhibit this low energy band[10] and we have shown that, in the case of thick ZnMgO shells on a ZnO core, the addition of Mg doesnot lead to the creation of new impurity centers (see Figure 1).The third contribution (noted (3) on Figure 3), with energies higher than the ZnO gap, is observed along the whole nanowire.We attribute this optical emission to quantum confinement in the radial wells. This contribution is broad(between 3.45 eV and 3.58 eV), which can be attributed to non-uniform thicknesses of the radial wells. Indeed, a close look at the TEM images in figure 2, reveals that the QW thickness varies from one m-plane radial well to the other and, further, is not strictly constant in the wells themselves. Moreover, m-planes and a-planes quantum wellshave different thicknesses. Finally, the thickness of the quantum wells is lower at the bottom of a nanowire than at the top, as evidenced by the inverse tapering shape of the nanowires visible on the SEM image of Figure 2 (a) and Figure 4.

**5. Correlation between structural and optical properties**

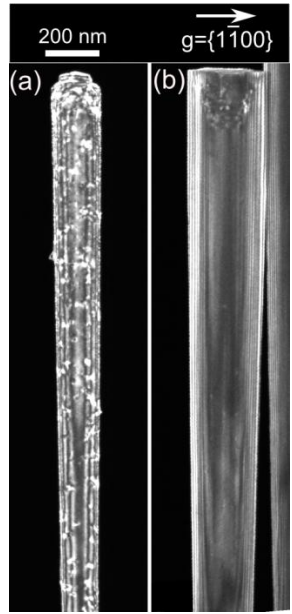

Figure 4: Weak-beam TEM images with g={1-100} diffracting conditions of (a) MQW2 (30% Mg) and (b) MQW1(15% Mg).

Weak-beam TEM was performed on MQW2 and MQW1 samples, to assess the influence of the ZnMgO composition on the generation of dislocations (Figure 4). In Figure 4 (a), the image of a MQW2 nanowire is presented. Many dislocations are visible. This is attributed to the relaxation of the lattice misfit between the ZnO core and the ZnMgO barrier. Indeed, the relative variation of $a$ lattice parameter $\Delta a / a$ is +0.6%, and the relative variation of $c$ lattice parameter (in the [0001] direction) $\Delta c / c$ is -0.7% for 30% Mg, according to Ohtomo *et al.*[25] For sample MQW1, nearly no dislocations are observed on the weak-beam TEM image [Figure 4 (b)]. Because the misfit between ZnO and ZnMgO is twice smaller in sample MQW1 than in sample MQW2, the ZnMgO barriers are elastically deformed on the ZnO core, and nearly no misfit dislocations are generated. To give a better insight of the misfit relaxation mechanism, the identification of the nature of the dislocation is presently being carried out.

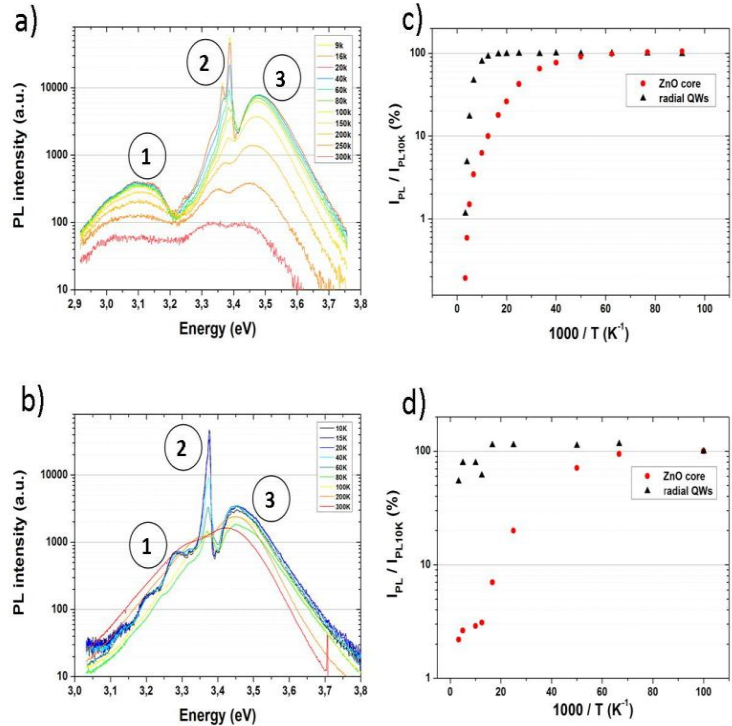

Figure 5 : Temperature dependent PL of (a) MQW2 sample (ZnMgO barriers with 30% Mg) and of (b) MQW1 sample (ZnMgO barriers with 15% Mg). PL intensity (normalized to 10K value) vs. 1000/T.for MQW2 (c) , and for MQW1 (d), for ZnO core (red dots) and for the radial wells (black triangles)

To follow the influence of the Mg composition in the ZnMgO barrier on the optical properties of the core-shell quantum wells, PL was recorded as a function of temperature from 10K to 300K for samples MQW2(a) and MQW1 (b)( Cf. figure 5).For MQW2 and MQW1, the three contributions as already observed by CL on single nanowires in Figure 3 are found on the low temperature PL spectra of the NW ensemble. The contributions from the radial wells(denoted 3 in

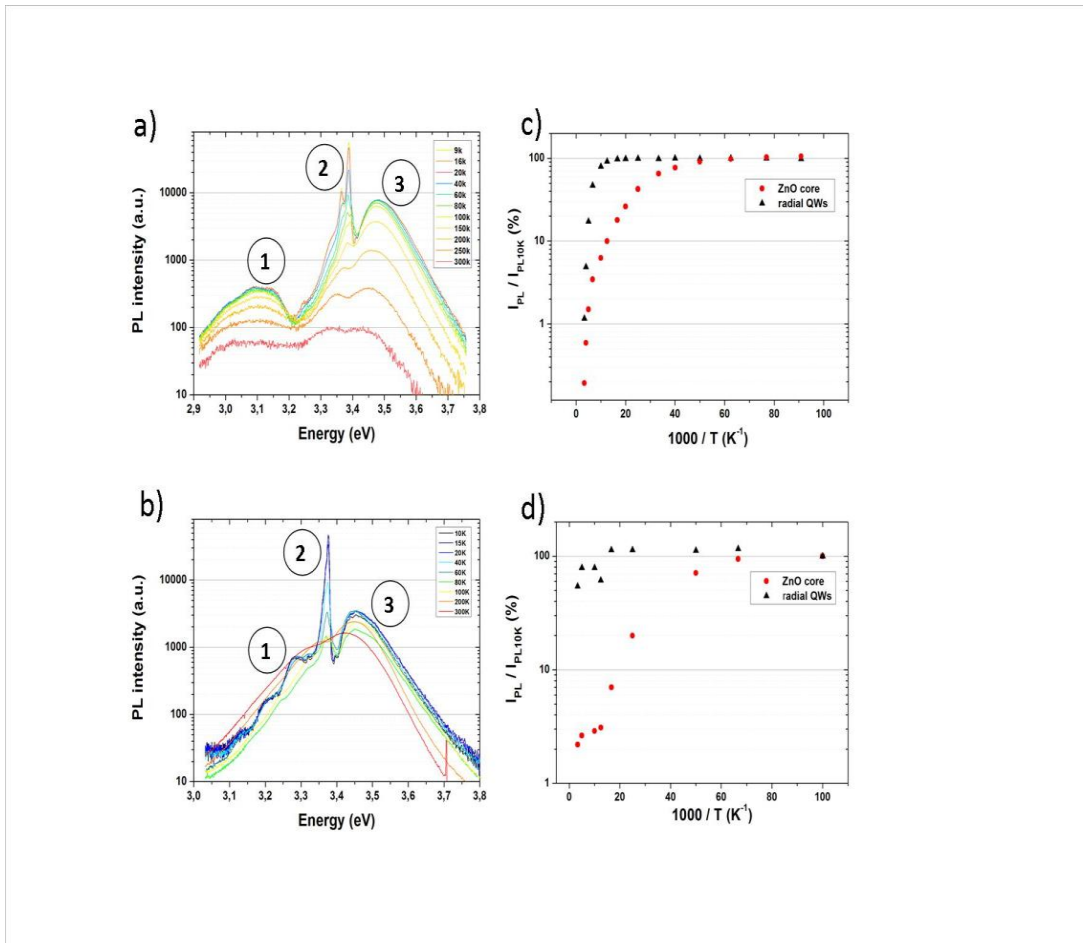

Figure 5) and axial wells(denoted 1) are broad: as discussed above, thickness variations are observed within the wells and also between different wells. Furthermore, the dispersion in ZnO NW densitiesmay induce radial growth rate dispersion among the nanowires. For both samples,as evidenced in fig.5(c) and 5(d), the intensity of the band edge emission from the ZnO core is rapidlyquenchedabove20K while the emission from the radial quantum wellskeeps constant up to 80-100K. This may be attributed to three-dimensionalconfinement into the wells possibly due to composition fluctuations in the barrier alloy or to thickness variations of the quantum wells:this would make it more difficult for confined excitons to recombine on non-radiative defects. For the higher Mg concentration sample MQW2, the quantum well PL IQE reaches only 1%. This is in contrast with the lower Mg concentration sample MQW1, where this ratio reaches the remarkably high value of 54%. This result is correlated with thepresence or the absence of dislocations, as observed on Figure 4. The Mg rich sample contains many dislocations [Figure 4(a)] which act as non-radiative recombination centers, explaining the very low emission at room temperature for the radial quantum wells. This is in contrast with the sample with low Mg concentration which does not contain any dislocations [Figure 4(b)]: in this case there are much less non-radiative centers and the emission of the radial quantum wells is maintained at a high level even at room temperature.

To summarize, vertical arrays of defect free nanowires with *m*-plane facets have been used as templates for growing radial ZnO/ZnMgO heterostructures with 3 ZnO quantum wells. Axial and radial quantum wells were evidenced by STEM HAADF. Their respective luminescence below and above band gap are determined by CL and explained by quantum confinement effect with QCSE in the axial wells. For barriers with 15 and 30% Mg compositions, a clear correlation is found between the structural properties as observed by TEM and the PL IQE. In Mg-rich core-shell heterostructures, misfit induced dislocationslead to a faster decrease of the optical emissionwith temperature. In contrast, for the lower Mg concentration, a PL IQE as high as 54% is obtained, in agreement with the absence of structural defects as observed by TEM. To the best of our knowledge, this value of PL IQE is a record value for wide band gap core-shell nanowires. Selective area growth is planned to obtain a better control of the nanowire density and sizes, with better emission properties in terms of intensity and spectral range. Moreover, it would be beneficial to resort to ZnMgO core instead of ZnO: in this case, recombinationswould occur only in the ZnO wells and misfit between the core and the ZnMgO barriers would be avoided, allowing for efficient optical emission atsmaller wavelength.

ACKNOWLEDGMENT We thank Eric Gautier for itshelp during the FIB sample preparation, and the French national research agency (ANR) for funding through the Carnot program (2006/2010).